\documentclass[apjl]{emulateapj}
\usepackage{amssymb}
\usepackage{graphicx,graphics}
\usepackage{epsfig}
\usepackage{dcolumn}
\usepackage{bm}
\usepackage{natbib}
\usepackage{times}
\usepackage{ulem}
\def\kmm#1  {{\bf [KMM:~ #1]~}}
\def\new#1 {{\bf #1 }}
\def\cut#1 {\sout{#1} }

\newcommand{\pmn}{PMN\,J0134$-$0931}
\newcommand{\pks}{PKS\,1413+135}

\newcommand{\dal}{\ensuremath{\lsb \Delta \alpha/ \alpha \rsb}}

\newcommand{\dmu}{\ensuremath{\lsb \Delta \mu/\mu \rsb}}
\newcommand{\beq}{\begin{equation}}
\newcommand{\eeq}{\end{equation}}

\newcommand{\lsb}{\left[}
\newcommand{\rsb}{\right]}
\newcommand{\hi}{H{\sc i}}
\newcommand{\kms}{km~s$^{-1}$}

\shorttitle{Constraining fundamental constant evolution}
\shortauthors{Kanekar et al.}
\begin{document}
\title{Constraining fundamental constant evolution with H{\sc i} and OH lines}

\author{N. Kanekar\altaffilmark{1},
G. I. Langston\altaffilmark{2},
J. T. Stocke\altaffilmark{3},
C. L. Carilli\altaffilmark{4},
K. L. Menten\altaffilmark{5}
}
\altaffiltext{1}{Ramanujan Fellow, National Centre for Radio Astrophysics, 
TIFR, Ganeshkhind, Pune - 411007, India; nkanekar@ncra.tifr.res.in}
\altaffiltext{2}{National Radio Astronomy Observatory, Green Bank, WV 24944, USA}
\altaffiltext{3}{CASA, Department of Astrophysical and Planetary Sciences, University of Colorado, 389-UCB, Boulder, CO 80309, USA}
\altaffiltext{4}{National Radio Astronomy Observatory, 1003 Lopezville Road, Socorro, NM87801, USA}
\altaffiltext{5}{Max-Planck-Institut f\"{u}r Radioastronomie, Auf dem H\"{u}gel 69, 53121, Bonn, Germany}

\begin{abstract}
We report deep Green Bank Telescope spectroscopy in the redshifted H{\sc i}\,21cm 
and OH\,18cm lines from the $z = 0.765$ absorption system towards PMN\,J0134$-$0931. 
A comparison between the ``satellite'' OH\,18cm line redshifts, or between the 
redshifts of the H{\sc i}\,21cm and ``main'' OH\,18cm lines, is sensitive to 
changes in different combinations of three fundamental constants, the fine 
structure constant $\alpha$, the proton-electron mass ratio $\mu \equiv m_p/m_e$
and the proton g-factor $g_p$. We find that the satellite OH\,18cm lines are not 
perfectly conjugate, with both different line shapes and stronger 1612~MHz absorption 
than 1720~MHz emission. This implies that the satellite lines of this absorber are not 
suitable to probe fundamental constant evolution. A comparison between the redshifts 
of the H{\sc i}\,21cm and OH\,18cm lines, via a multi-Gaussian fit, yields the strong 
constraint $\left[ \Delta F/F \right] = [-5.2 \pm 4.3] \times 10^{-6}$, where 
$F \equiv g_p \left[ \mu \alpha^2 \right]^{1.57}$ and the error budget includes 
contributions from both statistical and systematic errors. We thus find no 
evidence for a change in the constants between $z = 0.765$ and the present 
epoch. Incorporating the constraint $\left[ \Delta \mu/\mu \right] < 3.6 \times 10^{-7}$ 
from another absorber at a similar redshift and assuming that fractional changes 
in $g_p$ are much smaller than those in $\alpha$, we obtain 
$\left[ \Delta \alpha/\alpha \right] = (-1.7 \pm 1.4) \times 10^{-6}$ 
over a lookback time of 6.7~Gyrs.

\end{abstract}


\maketitle
\section{Introduction} 
\label{sec:intro}

The standard model of particle physics implicitly assumes that the values of coupling constants 
and particle masses do not depend on space or time. Conversely, variation in such ``fundamental
constants'' appears to be a generic feature of higher-dimensional theories aiming to unify the 
standard model and general relativity \citep[e.g.][]{marciano84}. Studies of fundamental constant 
evolution are hence of much interest as they both probe the foundations of the standard model and 
allow the possibility of distinguishing between different unification models at low energy scales 
\citep{uzan11}.


While laboratory atomic clock studies have yielded strong constraints on short-term changes in the 
fine structure constant $\alpha$ \citep[e.g.][]{rosenband08}, such studies are not sensitive to 
changes on Gyr timescales.  A wide range of methods, based on various spectral transitions, 
has been used to probe changes in $\alpha$, the proton-electron mass ratio $\mu \equiv m_p/m_e$ and 
the proton g-factor $g_p$ on cosmological timescales 
\citep[e.g.][]{bahcall67,thompson75,wolfe76b,varshalovich93,dzuba99,darling03,chengalur03,kanekar04a,flambaum07b}.  
Significant progress has recently been made in both the development of new techniques and 
the sensitivity of measurements. A combination of the many-multiplet method \citep{dzuba99} 
with spectra from the High Resolution Echelle Spectrograph (HIRES) on the Keck telescope has found
evidence for changes in $\alpha$ with redshift, $\dal = (-5.7 \pm 1.1) \times 10^{-6}$ 
for 143~absorbers at an average redshift ${\bar z} = 1.75$ \citep{murphy04}. Later studies 
applying this method to spectra from the Ultraviolet Echelle Spectrograph (UVES) on the Very Large 
Telescope (VLT) have not confirmed this result \citep[e.g.][]{srianand07b,molaro08}. Recently, 
\citet{webb11} applied the many-multiplet method to a large VLT-UVES sample and also did not find 
evidence supporting the Keck-HIRES result. While \citet{webb11} attempted to reconcile the 
Keck-HIRES and VLT-UVES results by proposing spatio-temporal changes in $\alpha$, a simpler 
explanation is that the errors in both studies have been under-estimated, especially given that 
systematic and unexplained errors have been shown to be present in the wavelength calibration 
of both spectrographs \citep{griest10,whitmore10,agafonova11}. Keck-HIRES and VLT-UVES spectra 
in redshifted H$_2$ lines have also yielded constraints on changes in $\mu$: the best current 
result is $\dmu < 4.4 \times 10^{-6}$ (${\bar z} \sim 2.8$; \citealp{king11}), although the 
errors here too may have been under-estimated due to wavelength calibration issues.

Radio spectroscopic techniques provide independent probes of fundamental constant evolution, with 
different systematic effects from those in optical schemes \citep[e.g.][]{kanekar08b}. Such methods include 
comparisons between rotational and \hi\,21cm hyperfine lines \citep{drinkwater98}, between different 
hydroxyl (OH) lines or \hi\,21cm and OH\,18cm lines \citep{darling03,chengalur03,kanekar04a}, between 
rotational and ammonia (NH$_3$) inversion lines \citep{flambaum07b}, between different methanol or hydronium 
lines \citep{jansen11,levshakov11,kozlov11}, between far-infrared fine structure lines and rotational lines 
\citep{levshakov08}, 
etc, each sensitive to different combinations of $\alpha$, $\mu$ and $g_p$. For example, the inversion-rotation 
comparison has yielded the best current constraint on changes in $\mu$ from {\it any} astronomical method, 
$\dmu < 3.5 \times 10^{-7}$ over $0 < z < 0.685$ \citep{kanekar11}. 

In general, techniques using multiple spectral lines from a single atomic or molecular species 
(e.g., OH, methanol, Fe{\sc ii}, etc) are preferable to those using different species, as
lines in the former case are likely to arise in the same gas, making the technique less susceptible 
to local velocity offsets. The satellite OH\,18cm lines have the special property of having exactly 
the same shape and opposite sign in certain astrophysical circumstances, due to a population inversion 
mechanism and quantum mechanical selection rules \citep{elitzur92,langevelde95}.  This ``conjugate'' 
behavior makes them ideal probes of fundamental constant evolution \citep{kanekar04b,kanekar08b}. 
Only two such conjugate OH\,18cm systems have so far been discovered at cosmological distances, at 
$z \sim 0.247$ towards \pks\ \citep{kanekar04b,darling04} and $z \sim 0.765$ towards \pmn\ 
\citep{kanekar05}. A high-sensitivity study of \pks\ with the Westerbork Synthesis Radio Telescope and 
the Arecibo Telescope found tentative evidence (at 99.1\% confidence level) for changes in $\alpha$, 
$\mu$ and/or $g_p$ \citep{kanekar10b}. In this {\it Letter}, we report deep Green Bank Telescope (GBT) 
observations of the redshifted \hi\,21cm and OH\,18cm lines in the $z \sim 0.765$ system towards \pmn, 
that yield strong constraints on changes in the fundamental constants.


\section{Observations, data analysis and spectra}
\label{sec:data}

\begin{figure*}[t!]
\includegraphics[scale=0.4]{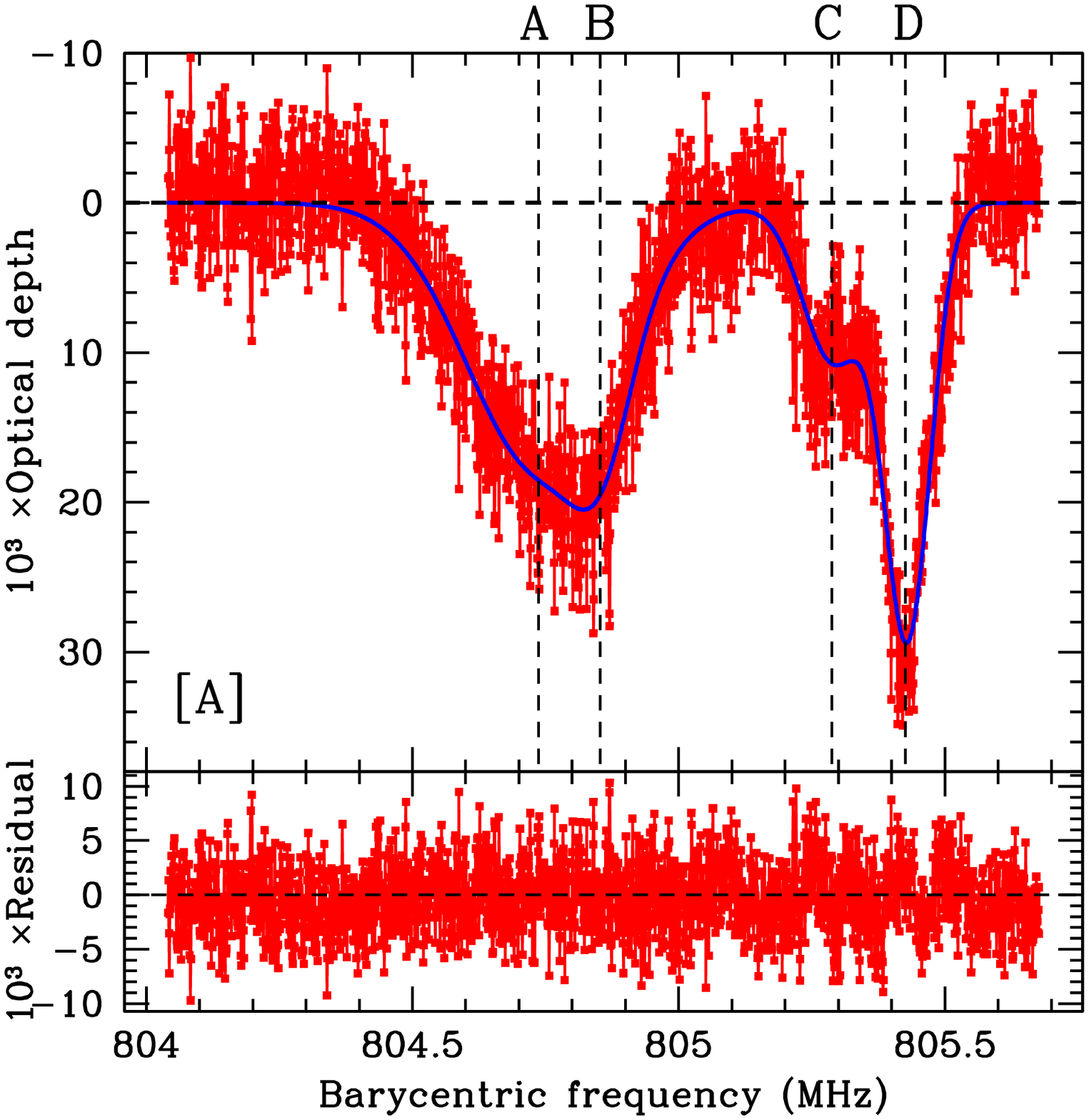}
\includegraphics[scale=0.4]{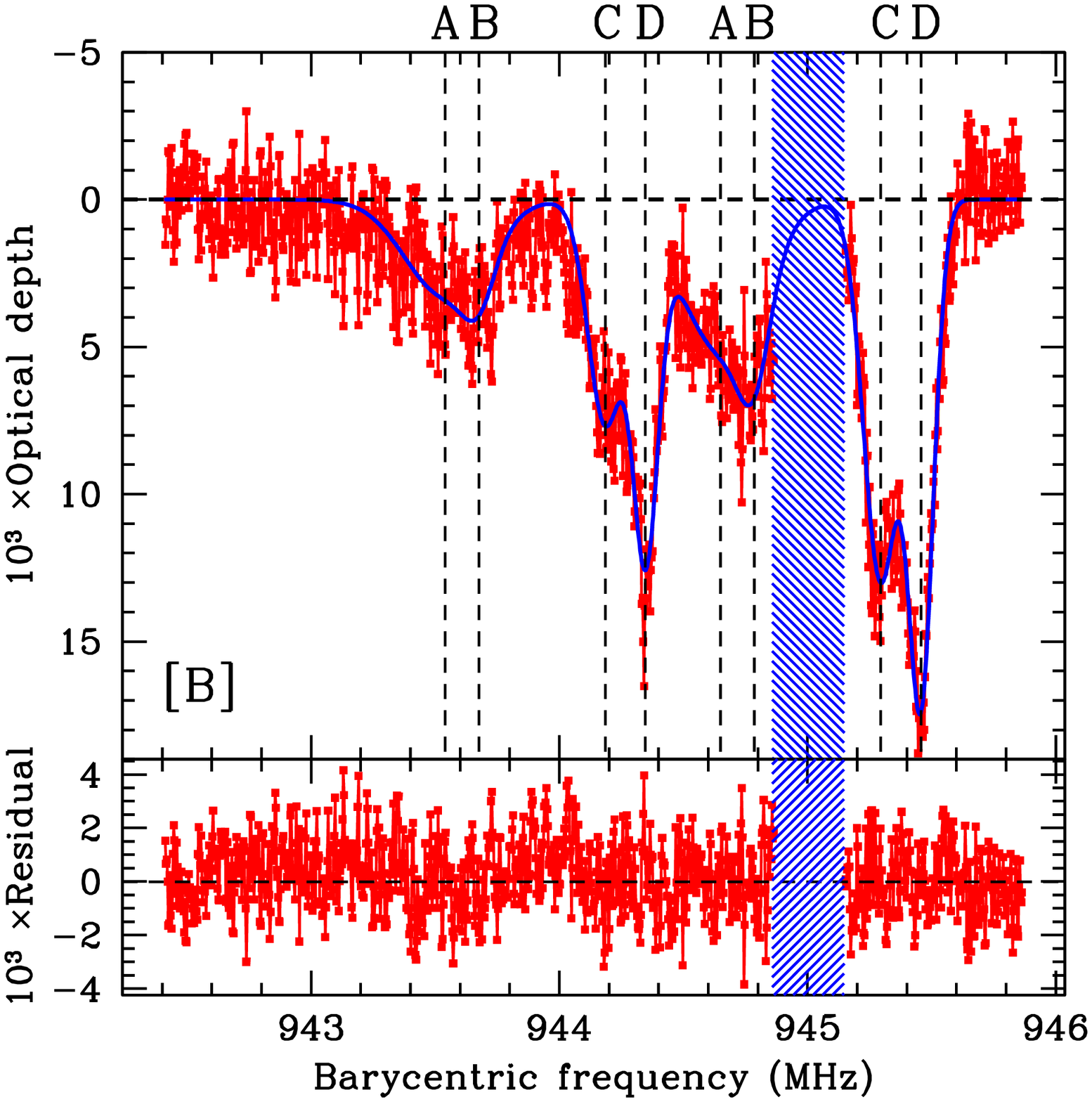}
\caption{GBT optical depth spectra in the redshifted [A]~\hi\,21cm and [B]~``main'' OH\,18cm
transitions; the solid line in each upper panel shows the 4-component fit, with the fit residuals 
shown in the lower panels. Shaded regions indicate frequencies affected by RFI. 
\label{fig:hi-oh}}
\end{figure*}

\begin{figure}[t!]
\includegraphics[scale=0.4]{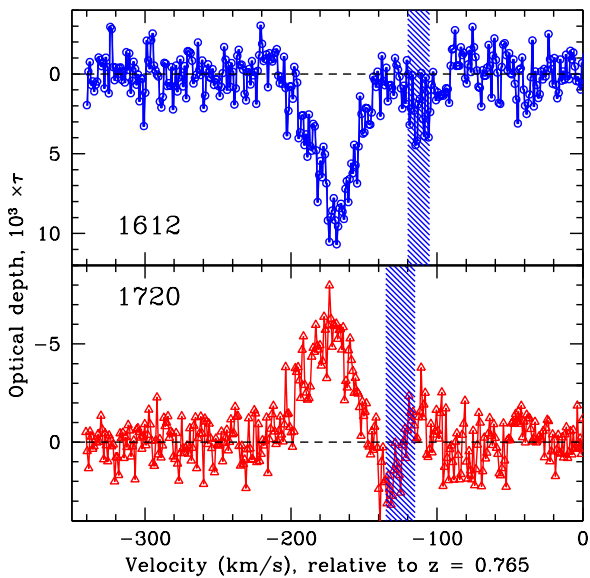}
\caption{GBT optical depth spectra in the redshifted OH\,1612 MHz line (upper panel) and 
OH\,1720~MHz line (lower panel). Shaded regions indicate frequencies affected by RFI.
\label{fig:ohsat}}
\end{figure}

The GBT observations of the \hi\,21cm and OH\,18cm lines from the $z \sim 0.765$ gravitational 
lens towards \pmn\ \citep{winn02,kanekar03d} were carried out between September 2005 and October 
2006 (proposal AGBT05C-037), using the PF1-800 and PF2 receivers, respectively. The observations 
used the AutoCorrelation Spectrometer (ACS) with 9-level sampling, two polarizations and 10-second 
integrations. A single ACS 12.5~MHz band, sub-divided into 32768~channels, was used for the \hi\,21cm 
line, while four ACS 12.5~MHz sub-bands, each sub-divided into 8192~channels, were used for the 
ground-state OH\,18cm lines. The system passband was calibrated by position-switching every five 
minutes, with system temperatures measured using a noise diode. The on-source times were 
$\sim 15$~hours for the \hi\,21cm line and $\sim 30$~hours for the OH\,18cm lines.

All data were analysed in the package {\sc DISH}, using standard procedures. The OH\,18cm data were
especially affected by intermittent radio-frequency interference (RFI); a visual inspection of 
every calibrated 10~second integration was used to excise data affected by RFI. Most of the data 
were found to have narrow-band RFI at $\sim 945$~MHz, close to the centre of the OH\,1667~MHz 
absorption profile. The channels affected by RFI were entirely edited out in the final spectrum, 
along with 25~channels on either side (i.e. the frequency range $944.867-945.150$~MHz). Similarly,
weak RFI was also found adjacent to the satellite OH\,18cm profiles, due to which a few channels 
in the final spectra were blanked out.

The final \hi\,21cm and ``main'' OH\,18cm optical depth spectra are shown in the upper panels of 
Fig.~\ref{fig:hi-oh}, while the ``satellite'' OH\,18cm spectra are shown in Fig.~\ref{fig:ohsat}. 
The \hi\,21cm and OH\,18cm spectra have velocity resolutions of $\sim 0.3$~km/s and $\sim 1.0$~km/s, 
respectively (after Hanning-smoothing and resampling), and root-mean-square optical depth noise values of 
$0.0019$ per 0.3~km/s channel (\hi\,21cm) and $\sim 0.0011-0.0012$ per 1.0~km/s channel (OH\,18cm lines). 
Note that these are ``apparent'' optical depths, derived using the total flux density of \pmn\ at the 
line frequencies. \pmn\ is a 5-component gravitational lens system with an angular extent of 
$\approx 0.7''$ \citep{winn02}, (i.e. $\approx 5$\,kpc at $z = 0.765$ and unresolved by the GBT beam). 
The \hi\,21cm and OH\,18cm features are likely to arise against only one or two of the source 
components, implying that the ``true'' optical depths are probably significantly larger than the 
measured optical depths.

Fig.~\ref{fig:ohsat} shows that, while the 1720~MHz line is in emission and the 1612~MHz line is 
in absorption, the two lines have {\it different} strengths, with the peak optical depths in the 
1720~MHz and 1612~MHz lines being $\approx -0.007$ and $\approx 0.01$, respectively.  
Thus, although the satellite OH lines have similar shapes, they are not exactly conjugate, 
with the 1612~MHz line about $1.5$ times stronger than the 1720~MHz line. We tried a number 
of analysis procedures and RFI excision schemes to test whether the difference between the satellite 
line profiles might arise due to RFI; the difference was found to be present in all cases. Further, 
the flux densities of \pmn\ measured in the 1612~MHz and 1720~MHz spectra were very similar ($\sim 0.7$~Jy \
in each), and the RMS noise values on the two spectra are comparable, indicating that there is no 
scaling error in the flux density calibration. It is thus unlikely that the difference between 
the satellite line profiles arises due to either RFI or problems with the flux density scale.

\section{Probing fundamental constant evolution}
\label{sec:alpha}

\subsection{The satellite OH lines}
\label{sec:ohsat}

The $2\Pi_{3/2} (J=3/2)$ OH rotational ground state is split into four sub-levels by
$\Lambda$-doubling and hyperfine splitting; two sub-levels have total angular momentum 
quantum number $F=2$, while the other two have $F=1$. The satellite OH\,18cm lines correspond 
to transitions with $\Delta F = \pm 1$ between these sub-levels. Similarly, two sub-levels of 
every excited rotational state have $F = J + 1/2$, while the other two have $F = J - 1/2$. 

The satellite OH\,18cm lines are said to be ``conjugate'' when they have the same shape, but with 
one line in emission and the other in absorption \citep{elitzur92}. This arises because, when the
OH molecules are pumped to excited states (by collisions or far-infrared radiation), the downward 
cascade to the ground state yields population inversion in the ground state sub-levels as certain 
transitions are forbidden by the selection rules $\Delta F = 0,\: \pm 1$. If the last stage of the 
cascade is the intra-ladder $119\mu$m transition $2\Pi_{3/2} (J=5/2) \rightarrow 2\Pi_{3/2} (J=3/2)$, 
transitions between the $F=3$ and $F=1$ sub-levels are forbidden; this would over-populate sub-levels 
with $F=2$ relative to those with $F=1$. As a result, the 1720~MHz transition would be inverted while 
the 1612~MHz transition would be anti-inverted; this is the situation in Fig.~\ref{fig:ohsat}, 
with the 1720~MHz and 1612~MHz transitions in emission and absorption, respectively.

If the $119\mu$m lines connecting the excited and ground states are optically thick \citep[which 
depends on the local particle number density and velocity gradient;][]{elitzur76,guibert78}, the 
rate coefficients for the different branches of the cascade are independent of line strength. This 
implies that the same number of particles are present in the two $F=2$ ground-state sub-levels, and, 
similarly, in the two $F=1$ sub-levels. The 1720~MHz and 1612~MHz lines then have identical 
strengths and shapes, albeit opposite sign \citep{elitzur92}. This is the ideal situation for the use 
of the satellite OH\,18cm lines to probe changes in the fundamental constants as the identical 
line shapes guarantee that the lines arise from the same gas \citep{kanekar04a,kanekar08b}.

Exactly conjugate satellite OH\,18cm lines have been observed in several extragalactic sources including 
Cen\,A \citep{langevelde95}, M82 \citep{seaquist97}, NGC\,253 \citep{frayer98}, 
and PKS\,1413+135 \citep{kanekar10b}. In Cen\,A and NGC\,253, the satellite lines are conjugate over 
a wide range of conditions and even show the cross-over from absorption to emission (and vice-versa) 
in each transition. As such, the fact that the satellite lines towards \pmn\ are {\it not} perfectly 
conjugate is both unusual and unexpected. If the $119\mu$m transitions that dominate the downward cascade 
are not optically thick, the rate coefficients of the decay routes to the different sub-levels are 
different and one would not obtain conjugate behavior \citep{elitzur76}. Alternatively, there may be 
absorption at the satellite line velocities from the molecular cloud that gives rise to part of 
the main OH\,18cm absorption. For example, there is a main OH\,18cm component at $z = 0.76385$ (see 
Table~1), with peak 1667~MHz opacity $\approx 0.0129$. If the OH ground-state levels in 
this cloud are thermalized, the satellite lines would have peak optical depths of $\approx 0.0014$ (nine 
times weaker than the 1667~MHz line). The 1612~MHz absorption would hence be increased, and the 
1720~MHz emission reduced, by this amount, yielding a difference of $\approx 0.0028$ between 
the satellite optical depths, consistent with the observed difference. 

Since the satellite OH\,18cm lines are not perfectly conjugate, it cannot be assumed that the lines arise 
from the same gas. The lines also have slightly different shapes, in addition to the different strengths. 
This implies unknown systematic effects if these lines are used in studies of fundamental constant 
evolution. We hence conclude that the satellite OH\,18cm lines in the $z \sim 0.765$ absorber 
towards \pmn\ are not suitable to probe changes in $\alpha$, $\mu$ and $g_p$.

\subsection{The ``main'' OH\,18cm and \hi\,21cm lines}
\label{sec:ohmain}

A comparison between the redshifts of the ``main'' OH\,18cm lines and the \hi\,21cm line is sensitive 
to changes in $F \equiv g_p \left[ \alpha^2 \mu \right]^{1.57}$ \citep{chengalur03}. We used a 
multi-Gaussian fit to the \hi\,21cm and main OH\,18cm lines to test for changes in $F$, assuming 
the same velocity structure in the different lines. Independent fits to the \hi\ and OH lines 
found a 4-component model to give a good fit to each line, with good agreement between the line widths 
of the corresponding \hi\ and OH components.  Turbulent broadening was hence assumed to dominate 
the line widths, with the \hi\ and OH widths of each component tied together in the fit. No 
assumption was made about the relative strengths of the main OH lines. However, since the main 
OH\,18cm line frequencies have the same dependence on $\alpha$, $\mu$ and $g_p$ to first order 
\citep{chengalur03,kozlov09}, the OH line redshifts of each component were tied together. The fit included a 
single velocity offset between all \hi\,21cm and OH\,18cm components, to account for the putative 
change in $F$. 

With the above assumptions, we used {\sc VPFIT} to carry out a simultaneous multi-Gaussian fit to 
the \hi\,21cm and OH\,18cm profiles, varying the fit parameters to minimize $\chi^2_\nu$. A four-component
model yielded $\chi^2_\nu = 1.09$ and noise-like residuals after subtracting out the fit. A Kolmogorov-Smirnov 
rank-1 test found the residual spectra (after subtracting out the fit) to be consistent (within $\sim 2\sigma$
significance) with a normal distribution. Fits with fewer than four components were found to yield 
a significantly higher $\chi^2_\nu$; conversely, increasing the number of components did not improve 
the fit. 

A four-component model thus provides a good fit to the \hi\,21cm and OH\,18cm profiles. The number of 
fitted parameters here is 21, four redshifts, four line widths, twelve peak line depths 
in the \hi\,21cm, OH\,1665 and OH\,1667 spectra, and the velocity offset between the \hi\,21cm and OH\,18cm 
lines. The parameters of the fit are summarized in Table~1; the error on each parameter has 
been increased by a factor $\sqrt{\chi^2_\nu}$, to account for the fact that the best-fit 
value of $\chi^2_\nu$ is slightly larger than unity ($\chi^2_\nu = 1.09$), probably because 
the RMS optical depth noise on the spectra has been marginally under-estimated. The best-fit velocity 
offset is $\Delta V = (-1.57 \pm 0.44)$\,\kms, with the \hi\,21cm line blueward of the OH\,18cm 
lines.

The above error on $\Delta V$ is the statistical error from the fit. Other contributions to the error 
budget include errors in the line rest frequencies and the frequency scale calibration, and local 
velocity offsets between the clouds giving rise to the \hi\,21cm and OH\,18cm lines; the latter
dominate the systematic errors. \citet{kanekar05} estimated the Galactic dispersion between 
\hi\,21cm and OH\,18cm velocities to be $< 1.2$\,\kms, using the measured dispersion between Galactic 
\hi\,21cm and HCO$^+$ velocities \citep{drinkwater98} and the good match between HCO$^+$ and OH velocities 
\citep{liszt00}. We will assume that this dispersion also applies to the $z \sim 0.765$ absorber. 
For comparison, the main OH\,18cm frequencies have been measured with an accuracy of $12$\,Hz 
($\sim 2$\,m/s; \citealp{hudson06}), while the error in the GBT frequency scale (mainly due to 
Doppler tracking) is $\lesssim 15$\,m/s. 

Our final result for the velocity offset between \hi\,21cm and OH\,18cm lines, including both statistical 
and systematic errors, is thus $\Delta V = [-1.57 \pm 0.44 (stat.) \pm 1.2 (syst.)]$\,\kms. This yields 
$\left[ \Delta F/F \right] = [-5.2 \pm 1.5 (stat.) \pm 4.0 (syst.)] \times 10^{-6}$. Adding the 
statistical and systematic errors in quadrature gives $\left[ \Delta F/F \right] = [-5.2 \pm 4.3] 
\times 10^{-6}$. We thus find no evidence for a change in $\alpha$, $\mu$ or $g_p$ between $z = 0.765$ 
and today, i.e. over a period of $6.7$~Gyrs.\footnote{We use a standard LCDM cosmology, with 
H$_0 = 70.4$\,\kms\,Mpc$^{-1}$, 
$\Omega_m = 0.27$ and $\Omega_\Lambda=0.73$ \citep{komatsu11}.}

\section{Discussion}
\label{sec:discussion}

\begin{figure}[t!]
\includegraphics[scale=0.4]{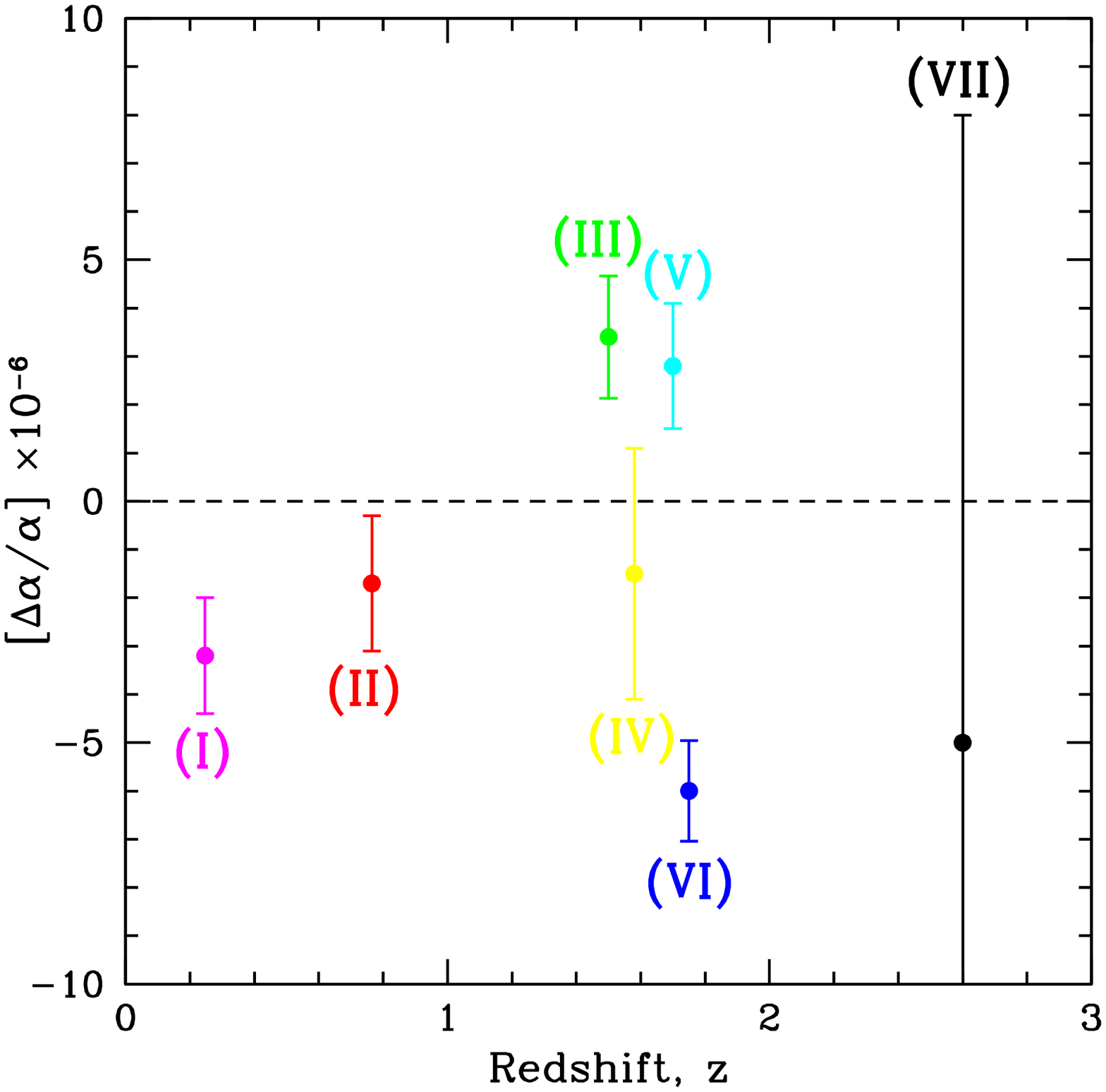}
\caption{A comparison between the best present estimates of $\dal$, from a variety of methods: 
(I)~conjugate satellite OH\,18cm \citep{kanekar10b}, (II)~\hi\,21cm and OH\,18cm lines (this work),
(III)~\hi\,21cm and C{\sc i} lines \citep{kanekar10}, (IV)~VLT-UVES many-multiplet 
\citep{agafonova11}, (V)~VLT-UVES many-multiplet \citep{webb11}, (VI)~Keck-HIRES many-multiplet 
\citep{murphy04} and (VII)~alkali doublet \citep{murphy01a}. Methods (I), 
(II) and (III) assume $\dmu << \dal$ and $\left[ \Delta g_p/g_p \right] << \dal$. Results from 
comparisons between the strongest or nearest metal and \hi-21cm components 
\citep[e.g.][]{tzanavaris05,srianand10} have been excluded as they have unknown 
systematic errors \citep[e.g.][]{kanekar06,kanekar07}.
\label{fig:dal}}
\end{figure}

The previous section constrains changes in $F \equiv g_p \left[ \alpha^2 \mu \right]^{1.57}$ 
and does not yield independent constraints on changes in $\alpha$, $\mu$ and $g_p$. However, 
this result can be used to obtain the sensitivity to changes in each parameter by assuming 
that the other two do not vary with time. The $1\sigma$ sensitivities are 
$\dal = 1.4 \times 10^{-6}$, $\dmu = 2.7 \times 10^{-6}$ and 
$\left[ \Delta g_p/g_p \right] = 4.3 \times 10^{-6}$.  A stringent constraint on changes 
in $\mu$ was recently obtained by \citet{kanekar11} from a comparison between NH$_3$ 
inversion and CS/H$_2$CO rotational lines at $z = 0.685$ towards B0218+357: 
$\dmu < 3.6 \times 10^{-7}$ between $z = 0.685$ and the present epoch. Assuming that there are 
no spatial variations in the constants, we can replace $\dmu$ in the expression for 
$\left[ \Delta F/F \right]$ to get $\left[ \Delta g_p/g_p \right] + 3.14 \times \dal 
= (-5.2 \pm 4.3) \times 10^{-6}$. If we further assume that $\left[ \Delta g_p/g_p \right] << \dal$,
we obtain $\dal = (-1.7 \pm 1.4) \times 10^{-6}$ between $z = 0.765$ and today. 

Fig.~\ref{fig:dal} shows a comparison between the best estimates of $\dal$ from a variety of 
techniques today \citep{murphy01a,murphy04,kanekar10b,kanekar10,webb11,agafonova11}. The Keck-HIRES 
many-multiplet dataset is the only one that finds statistically-significant evidence for changes in 
$\alpha$. However, systematic wavelength calibration errors may have been under-estimated here 
\citep{griest10}. At present, there appears to be no strong evidence from astronomical 
spectroscopy for changes in the fundamental constants on cosmological timescales.

In summary, we have carried out deep GBT spectroscopy in the redshifted \hi\,21cm and 
OH\,18cm transitions from the $z = 0.765$ absorber towards \pmn. We find that the 
satellite OH\,18cm lines are not perfectly conjugate, with the 1612~MHz
absorption $\sim 1.5$ times stronger than the 1720~MHz emission. The fact that 
the satellite lines have different shapes implies that these should not be used to probe 
fundamental constant evolution, due to the possibility of unknown systematic effects. We obtain 
tight constraints on changes in the quantity $F \equiv g_p \left[ \mu \alpha^2 \right]^{1.57}$ 
via a simultaneous fit to the \hi\,21cm and OH\,18cm line profiles. A four-component Gaussian model 
assuming turbulent line broadening yields a good fit to both profiles, yielding
$\left[ \Delta F/F \right] = [-5.2 \pm 4.3] \times 10^{-6}$, including both statistical 
and systematic errors. We find no evidence for a change in $\alpha$, $\mu$ or $g_p$ between 
$z = 0.765$ and today.

\setcounter{table}{0}
\begin{table*}
\label{table:fit} 
\begin{centering}
\begin{tabular}{|c|c|l|c|c|c|c|}
\hline
&&&&\multicolumn{3}{c|}{} \\ 
& Velocity offset &  Heliocentric redshift & FWHM & \multicolumn{3}{c|}{Optical depth $\times 100$} \\
&    km/s         &           &  km/s  &  \hi\,21cm & OH-1667 & OH-1665 \\
\hline
&&&&&&\\
A & $(-1.57 \pm 0.44)$  & $0.765057~~(11)$  & $120.5~ \pm 2.2~$ & $1.795 \pm 0.055$ & $0.509 \pm 0.024$ & $0.325 \pm 0.019$ \\
B & 			& $0.7648025 (78)$  & $~47.7~ \pm 4.3~$ & $0.67~ \pm 0.12~$ & $0.321 \pm 0.049$ & $0.167 \pm 0.035$ \\
C & 			& $0.7638501 (42)$  & $~51.9~ \pm 1.6~$ & $1.064 \pm 0.043$ & $1.291 \pm 0.038$ & $0.754 \pm 0.028$ \\
D &                     & $0.7635476 (25)$  & $~37.76 \pm 0.63$ & $2.868 \pm 0.055$ & $1.656 \pm 0.046$ & $1.119 \pm 0.035$ \\
&&&&&& \\
\hline
\end{tabular}
\caption{The parameters of the best 4-Gaussian fit  to the \hi\,21cm and OH\,18cm profiles; 
the second column contains the best-fit velocity offset between the lines.}
\end{centering}
\end{table*}

\acknowledgments
We thank Bob Carswell, Carl Bignell and Bob Garwood for much help with {\sc VPFIT}, and the GBT 
observations and data analysis. NK acknowledges support from the Department of Science and 
Technology, India, via a Ramanujan Fellowship. CC acknowledges 
support from the Max-Planck Society and the Alexander von Humboldt Foundation. JTS acknowledges 
support from NSF grant AST-0707480 and an NRAO travel grant. The National Radio Astronomy 
Observatory is operated by Associated Universities, Inc, under cooperative agreement with the NSF. 

\bibliographystyle{apj}
\bibliography{ms}

\begin{thebibliography}{49}
\expandafter\ifx\csname natexlab\endcsname\relax\def\natexlab#1{#1}\fi

\bibitem[{{Agafonova} {et~al.}(2011){Agafonova}, {Molaro}, {Levshakov}, \&
  {Hou}}]{agafonova11}
{Agafonova}, I.~I., {Molaro}, P., {Levshakov}, S.~A., \& {Hou}, J.~L. 2011,
  A\&A, 529, 28

\bibitem[{{Bahcall} {et~al.}(1967){Bahcall}, {Sargent}, \&
  {Schmidt}}]{bahcall67}
{Bahcall}, J.~N., {Sargent}, W.~L.~W., \& {Schmidt}, M. 1967, ApJ, 149, L11

\bibitem[{Chengalur \& Kanekar(2003)}]{chengalur03}
Chengalur, J.~N. \& Kanekar, N. 2003, Phys.~Rev.~Lett., 91, 241302

\bibitem[{Darling(2003)}]{darling03}
Darling, J. 2003, Phys.~Rev.~Lett., 91, 011301

\bibitem[{Darling(2004)}]{darling04}
---. 2004, ApJ, 612, 58

\bibitem[{{Drinkwater} {et~al.}(1998){Drinkwater}, {Webb}, {Barrow}, \&
  {Flambaum}}]{drinkwater98}
{Drinkwater}, M.~J., {Webb}, J.~K., {Barrow}, J.~D., \& {Flambaum}, V.~V. 1998,
  MNRAS, 295, 457

\bibitem[{Dzuba {et~al.}(1999)Dzuba, Flambaum, \& Webb}]{dzuba99}
Dzuba, V.~A., Flambaum, V.~V., \& Webb, J.~K. 1999, Phys.~Rev.~Lett., 82, 888

\bibitem[{Elitzur(1976)}]{elitzur76}
Elitzur, M. 1976, ApJ, 203, 124

\bibitem[{{Elitzur}(1992)}]{elitzur92}
{Elitzur}, M. 1992, {Astronomical Masers} (Dordrect, NL: Kluwer Academic)

\bibitem[{{Flambaum} \& {Kozlov}(2007)}]{flambaum07b}
{Flambaum}, V.~V. \& {Kozlov}, M.~G. 2007, Phys.~Rev.~Lett., 98, 240801

\bibitem[{{Frayer} {et~al.}(1998){Frayer}, {Seaquist}, \& {Frail}}]{frayer98}
{Frayer}, D.~T., {Seaquist}, E.~R., \& {Frail}, D.~A. 1998, AJ, 115, 559

\bibitem[{{Griest} {et~al.}(2010){Griest}, {Whitmore}, {Wolfe}, {Prochaska},
  {Howk}, \& {Marcy}}]{griest10}
{Griest}, K., {Whitmore}, J.~B., {Wolfe}, A.~M., {Prochaska}, J.~X., {Howk},
  J.~C., \& {Marcy}, G.~W. 2010, ApJ, 708, 158

\bibitem[{Guibert {et~al.}(1978)Guibert, Elitzur, \& N.-Q.-Rieu}]{guibert78}
Guibert, J., Elitzur, M., \& N.-Q.-Rieu. 1978, A\&A, 66, 395

\bibitem[{{Hudson} {et~al.}(2006){Hudson}, {Lewandowski}, {Sawyer}, \&
  {Ye}}]{hudson06}
{Hudson}, E.~R., {Lewandowski}, H.~J., {Sawyer}, B.~C., \& {Ye}, J. 2006,
  Phys.~Rev.~Lett., 96, 143004

\bibitem[{{Jansen} {et~al.}(2011){Jansen}, {Xu}, {Kleiner}, {Ubachs}, \&
  {Bethlem}}]{jansen11}
{Jansen}, P., {Xu}, L.-H., {Kleiner}, I., {Ubachs}, W., \& {Bethlem}, H.~L.
  2011, Phys. Rev. Lett., 106, 100801

\bibitem[{{Kanekar}(2008)}]{kanekar08b}
{Kanekar}, N. 2008, Modern Physics Letters A, 23, 2711

\bibitem[{{Kanekar}(2011)}]{kanekar11}
---. 2011, ApJ, 728, L12

\bibitem[{{Kanekar} \& {Briggs}(2003)}]{kanekar03d}
{Kanekar}, N. \& {Briggs}, F.~H. 2003, A\&A, 412, L29

\bibitem[{{Kanekar} {et~al.}(2005){Kanekar}, {Carilli}, {Langston}, {Rocha},
  {Combes}, {Subrahmanyan}, {Stocke}, {Menten}, {Briggs}, \&
  {Wiklind}}]{kanekar05}
{Kanekar}, N., {Carilli}, C.~L., {Langston}, G.~I., {Rocha}, G., {Combes}, F.,
  {Subrahmanyan}, R., {Stocke}, J.~T., {Menten}, K.~M., {Briggs}, F.~H., \&
  {Wiklind}, T. 2005, Phys. Rev. Lett., 95, 261301

\bibitem[{{Kanekar} \& {Chengalur}(2004)}]{kanekar04a}
{Kanekar}, N. \& {Chengalur}, J.~N. 2004, MNRAS, 350, L17

\bibitem[{{Kanekar} {et~al.}(2004){Kanekar}, {Chengalur}, \&
  {Ghosh}}]{kanekar04b}
{Kanekar}, N., {Chengalur}, J.~N., \& {Ghosh}, T. 2004, Physical Review
  Letters, 93, 051302

\bibitem[{{Kanekar} {et~al.}(2010{\natexlab{a}}){Kanekar}, {Chengalur}, \&
  {Ghosh}}]{kanekar10b}
---. 2010{\natexlab{a}}, ApJ, 716, L23

\bibitem[{{Kanekar} {et~al.}(2007){Kanekar}, {Chengalur}, \&
  {Lane}}]{kanekar07}
{Kanekar}, N., {Chengalur}, J.~N., \& {Lane}, W.~M. 2007, MNRAS, 375, 1528

\bibitem[{{Kanekar} {et~al.}(2010{\natexlab{b}}){Kanekar}, {Prochaska},
  {Ellison}, \& {Chengalur}}]{kanekar10}
{Kanekar}, N., {Prochaska}, J.~X., {Ellison}, S.~L., \& {Chengalur}, J.~N.
  2010{\natexlab{b}}, ApJ, 712, L148

\bibitem[{{Kanekar} {et~al.}(2006){Kanekar}, {Subrahmanyan}, {Ellison}, {Lane},
  \& {Chengalur}}]{kanekar06}
{Kanekar}, N., {Subrahmanyan}, R., {Ellison}, S.~L., {Lane}, W.~M., \&
  {Chengalur}, J.~N. 2006, MNRAS, 370, L46

\bibitem[{{King} {et~al.}(2011){King}, {Murphy}, {Ubachs}, \& {Webb}}]{king11}
{King}, J.~A., {Murphy}, M.~T., {Ubachs}, W., \& {Webb}, J.~K. 2011, MNRAS,
  417, 301

\bibitem[{{Komatsu} {et~al.}(2011){Komatsu}, {Smith}, {Dunkley}, {Bennett},
  {Gold}, {Hinshaw}, {Jarosik}, {Larson}, {Nolta}, {Page}, {Spergel},
  {Halpern}, {Hill}, {Kogut}, {Limon}, {Meyer}, {Odegard}, {Tucker}, {Weiland},
  {Wollack}, \& {Wright}}]{komatsu11}
{Komatsu}, E., {Smith}, K.~M., {Dunkley}, J., {Bennett}, C.~L., {Gold}, B.,
  {Hinshaw}, G., {Jarosik}, N., {Larson}, D., {Nolta}, M.~R., {Page}, L.,
  {Spergel}, D.~N., {Halpern}, M., {Hill}, R.~S., {Kogut}, A., {Limon}, M.,
  {Meyer}, S.~S., {Odegard}, N., {Tucker}, G.~S., {Weiland}, J.~L., {Wollack},
  E., \& {Wright}, E.~L. 2011, ApJS, 192, 18

\bibitem[{{Kozlov}(2009)}]{kozlov09}
{Kozlov}, M.~G. 2009, Phys. Rev. A, 80, 022118

\bibitem[{{Kozlov} \& {Levshakov}(2011)}]{kozlov11}
{Kozlov}, M.~G. \& {Levshakov}, S.~A. 2011, ApJ, 726, 65

\bibitem[{{Levshakov} {et~al.}(2011){Levshakov}, {Kozlov}, \&
  {Reimers}}]{levshakov11}
{Levshakov}, S.~A., {Kozlov}, M.~G., \& {Reimers}, D. 2011, ApJ, 738, 26

\bibitem[{{Levshakov} {et~al.}(2008){Levshakov}, {Reimers}, {Kozlov}, {Porsev},
  \& {Molaro}}]{levshakov08}
{Levshakov}, S.~A., {Reimers}, D., {Kozlov}, M.~G., {Porsev}, S.~G., \&
  {Molaro}, P. 2008, A\&A, 479, 719

\bibitem[{{Liszt} \& {Lucas}(2000)}]{liszt00}
{Liszt}, H. \& {Lucas}, R. 2000, A\&A, 355, 333

\bibitem[{{Marciano}(1984)}]{marciano84}
{Marciano}, W.~J. 1984, Phys. Rev. Lett., 52, 489

\bibitem[{{Molaro} {et~al.}(2008){Molaro}, {Reimers}, {Agafonova}, \&
  {Levshakov}}]{molaro08}
{Molaro}, P., {Reimers}, D., {Agafonova}, I.~I., \& {Levshakov}, S.~A. 2008,
  European Physical Journal Special Topics, 163, 173

\bibitem[{{Murphy} {et~al.}(2004){Murphy}, {Flambaum}, {Webb}, {Dzuba},
  {Prochaska}, \& {Wolfe}}]{murphy04}
{Murphy}, M.~T., {Flambaum}, V.~V., {Webb}, J.~K., {Dzuba}, V.~V., {Prochaska},
  J.~X., \& {Wolfe}, A.~M. 2004, in Lecture Notes in Physics, Vol. 648,
  Astrophysics, Clocks and Fundamental Constants, ed. S.~G. {Karshenboim} \&
  E.~{Peik} (Berlin: Springer-Verlag), 131

\bibitem[{{Murphy} {et~al.}(2001){Murphy}, {Webb}, {Flambaum}, {Dzuba},
  {Churchill}, {Prochaska}, {Barrow}, \& {Wolfe}}]{murphy01a}
{Murphy}, M.~T., {Webb}, J.~K., {Flambaum}, V.~V., {Dzuba}, V.~A., {Churchill},
  C.~W., {Prochaska}, J.~X., {Barrow}, J.~D., \& {Wolfe}, A.~M. 2001, MNRAS,
  327, 1208

\bibitem[{{Rosenband} {et~al.}(2008){Rosenband}, {Hume}, {Schmidt}, {Chou},
  {Brusch}, {Lorini}, {Oskay}, {Drullinger}, {Fortier}, {Stalnaker}, {Diddams},
  {Swann}, {Newbury}, {Itano}, {Wineland}, \& {Bergquist}}]{rosenband08}
{Rosenband}, T., {Hume}, D.~B., {Schmidt}, P.~O., {Chou}, C.~W., {Brusch}, A.,
  {Lorini}, L., {Oskay}, W.~H., {Drullinger}, R.~E., {Fortier}, T.~M.,
  {Stalnaker}, J.~E., {Diddams}, S.~A., {Swann}, W.~C., {Newbury}, N.~R.,
  {Itano}, W.~M., {Wineland}, D.~J., \& {Bergquist}, J.~C. 2008, Science, 319,
  1808

\bibitem[{{Seaquist} {et~al.}(1997){Seaquist}, {Frayer}, \&
  {Frail}}]{seaquist97}
{Seaquist}, E.~R., {Frayer}, D.~T., \& {Frail}, D.~A. 1997, ApJ, 487, L131

\bibitem[{{Srianand} {et~al.}(2007){Srianand}, {Chand}, {Petitjean}, \&
  {Aracil}}]{srianand07b}
{Srianand}, R., {Chand}, H., {Petitjean}, P., \& {Aracil}, B. 2007, Physical
  Review Letters, 99, 239002

\bibitem[{{Srianand} {et~al.}(2010){Srianand}, {Gupta}, {Petitjean},
  {Noterdaeme}, \& {Ledoux}}]{srianand10}
{Srianand}, R., {Gupta}, N., {Petitjean}, P., {Noterdaeme}, P., \& {Ledoux}, C.
  2010, MNRAS, 405, 1888

\bibitem[{{Thompson}(1975)}]{thompson75}
{Thompson}, R.~I. 1975, ApL, 16, 3

\bibitem[{Tzanavaris {et~al.}(2005)Tzanavaris, Webb, Murphy, Flambaum, \&
  Curran}]{tzanavaris05}
Tzanavaris, P., Webb, J.~K., Murphy, M.~T., Flambaum, V.~V., \& Curran, S.~J.
  2005, Phys.~Rev.~Lett., 95, 1301

\bibitem[{Uzan(2011)}]{uzan11}
Uzan, J.-P. 2011, Living Reviews in Relativity, 14, 2

\bibitem[{{van Langevelde} {et~al.}(1995){van Langevelde}, van Dishoek,
  Sevenster, \& Israel}]{langevelde95}
{van Langevelde}, H.~J., van Dishoek, E.~F., Sevenster, M.~N., \& Israel, F.~P.
  1995, ApJ, 448, L123

\bibitem[{{Varshalovich} \& {Levshakov}(1993)}]{varshalovich93}
{Varshalovich}, D.~A. \& {Levshakov}, S.~A. 1993, JETP, 58, L237

\bibitem[{{Webb} {et~al.}(2011){Webb}, {King}, {Murphy}, {Flambaum},
  {Carswell}, \& {Bainbridge}}]{webb11}
{Webb}, J.~K., {King}, J.~A., {Murphy}, M.~T., {Flambaum}, V.~V., {Carswell},
  R.~F., \& {Bainbridge}, M.~B. 2011, Phys. Rev. Lett., 107, 191101

\bibitem[{{Whitmore} {et~al.}(2010){Whitmore}, {Murphy}, \&
  {Griest}}]{whitmore10}
{Whitmore}, J.~B., {Murphy}, M.~T., \& {Griest}, K. 2010, ApJ, 723, 89

\bibitem[{{Winn} {et~al.}(2002){Winn}, {Lovell}, {Chen}, {Fletcher}, {Hewitt},
  {Patnaik}, \& {Schechter}}]{winn02}
{Winn}, J.~N., {Lovell}, J.~E.~J., {Chen}, H., {Fletcher}, A.~B., {Hewitt},
  J.~N., {Patnaik}, A.~R., \& {Schechter}, P.~L. 2002, ApJ, 564, 143

\bibitem[{{Wolfe} {et~al.}(1976){Wolfe}, {Brown}, \& {Roberts}}]{wolfe76b}
{Wolfe}, A.~M., {Brown}, R.~L., \& {Roberts}, M.~S. 1976, Phys. Rev. Lett., 37,
  179

\end{thebibliography}
\end{document}